# CCD Photometry of Variable Stars in the Globular Cluster NGC 288 [1]

J. Kaluzny

Warsaw University Observatory, Al. Ujazdowskie 4, 00-478 Warsaw, Poland

e-mail: @sirius.astrouw.edu.pl

## ABSTRACT

The central part of the globular cluster NGC 288 was surveyed in a search for short-period variable stars. We obtained V-band light curves for 4 SX Phe stars and 2 RR Lyrae variables, five of which are new discoveries. All SX Phe stars belong to blue stragglers. We present a color-magnitude diagram including stars from the tip of red giant branch to the level of the main-sequence turnoff. Three likely hot subdwarfs were located on the extension of the blue horizontal branch of the cluster.

## 1. Introduction

NGC 288 is a globular cluster of intermediate metallicity that is relatively easy to study owing to its proximity, low central concentration, and low reddening. It has been the subject of a number of photometric studies in the last decade (Bolte (1992) and references therein). The underlying motivation for most of these studies has been to find an explanation for the anomalously blue horizontal branch (HB) seen in the color-magnitude diagram (CMD) of this cluster. The most recent studies conclude that NGC 288 is older by a few billion years than other clusters of similar metallicity (Chaboyer et al. 1995; Sarajedini & Demarque (1990); VandenBerg et al. (1990)).

Only two variable stars were known in NGC 288 prior to our survey. Hoag (1973) catalogue lists the long period variable V1 discovered originally by Oosterhoff (1943). Hollingsworth and Liller (1977) identified an RR Lyr star in the cluster and named it V2.
NGC 288 harbors a large number of blue stragglers (Bolte 1992). Recent studies uncovered numerous short-period variables among blue stragglers in stellar clusters (Mateo 1995). Two classes of variables are observed: eclipsing binaries and pulsating stars of SX Phe type. The relative frequency of variable blue stragglers varies significantly from cluster to cluster. For example about 30% of blue stragglers in NGC5053 are short-period variables (Nemec et al. 1995) while no single variable star could be identified among about 27 BS in NGC 6366 (Harris 1993). It is not clear for a moment how the relative frequency of variable blue stragglers correlates with such parameters as a cluster age and metallicity. For example, Harris (1993) suggests that hight metallicity objects tend to have low frequency of SX Phe stars. Moreover, McNamara (1995) notes

---

that mean periods of SX Phe stars correlate with the metallicity of a parent cluster. In fact very few clusters were surveyed for variables located below the horizontal branch. In this contribution we present results of a preliminary survey in NGC 288.

## 2. Observations and data reduction

Our survey for variables in NGC 288 was conducted as a supplementary project during a long observing run devoted mainly to a search for eclipsing binaries in the globular cluster M4 (Kaluzny 1996). The central part of NGC 288 was monitored with the CTIO 0.9-m telescope and Tektronix 2048 No. 3 CCD. The field of view of the camera was $13.6 \times 13.6$ arcmin$^2$ with scale of 0.396 arcsec/pixel. The monitoring was performed in the Johnson V-band. A total of 81 useful V-band images were collected on 8 nights during the period July 6-18 (UT) 1995. The exposure time ranged from 240 to 600 seconds depending on the observing conditions. Most of frames were collected during dark time and with seeing better than 1.5 arcsec. In addition to observations in the V-band we took two long and one short exposures in the B-band. A table with a detailed log of all observations was submitted to editors of A&A (see Appendix A).

The preliminary processing of the raw data was made with IRAF [2]. The flat-field frames were prepared by combining sets of 10-15 frames obtained by observing an illuminated screen in the dome. The reduction procedures reduced total instrumental systematics to below 1% for the central $1500 \times 1500$ pixels$^2$ area of the images. Some systematic residual pattern at the 1%-2% level was left near borders of the images. This residual pattern had little effect on photometry presented below as most of measured stars (including all NGC 288 variables and all photometric standards from Landolt (1993) fields) were located in the central section of the images.

Stellar profile photometry was extracted using DoPHOT (Schechter et al. 1993). We used DoPHOT in the fixed-position mode. The stellar positions were provided based on a list obtained by reduction of a "template" image. The best of available images served as a "template". The images collected with the CTIO 0.9-m telescope suffer from a significant positional dependence of the point spread function. To cope with this effect we applied a procedure described in details in Kaluzny et al. (1996). In short, each analyzed frame was divided into a $5 \times 5$ grid of overlapping sub-frames. An instrumental photometry derived for a give sub-frame of a given frame was transformed to the common instrumental system of the "template" image. A data base containing measurements for a total of 16549 stars was constructed and subsequently used to select potentially variable stars. In Fig. 1 we present a plot of $rms$ deviation versus the average V magnitude for the light curves of 7098 stars which were measured on at least 45 frames. Only stars with median values of V magnitude lower than 20 were analysed for variability.

---

[2]IRAF is distributed by the National Optical Astronomy Observatories, which are operated by the Associations of Universities for Research in Astronomy, Inc., under cooperative agreement with the National Science Foundation

## 2.1. The color-magnitude diagram

Photometry derived from frames listed in Table 1 was used to construct the color-magnitude diagram (CMD) for the monitored field. The instrumental photometry was extracted using DAOPHOT/ALLSTAR (Stetson 1987). Following Walker (1994) we selected a Moffat-function point spread function, quadratically varying with X and Y coordinates. The instrumental photometry was transformed to the standard BV system using relations:

$$v = 2.9742 + V - 0.0102 \times (B - V) + 0.15 \times X \quad (1)$$
$$b - v = 0.3092 + 1.1376 \times (B - V) + 0.12 \times X \quad (2)$$

where X is the airmass and a lower case letters correspond to the instrumental magnitudes. This transformation was determined using photometry for 9 stars from two Landolt (1993) fields. Specifically, PG2213-006 and T Phe fields were observed at airmasses $X = 1.35$ and $X = 1.04$, respectively. For each field two exposures in the V-filter and two exposures in the B-filter were collected. The average extinction coefficients for CTIO were adopted. In Fig. 2 we show residuals between the standard and recovered magnitudes and colors for Landolt standards. Immediately after observations of standards we obtained frames of NGC 288 used for construction of the cluster CMD. Note that both fields with standards as well as NGC 288 were observed at small airmasses. Therefore, an uncertainty of the adopted extinction coefficients should have a little effect on the zero points of derived BV photometry of NGC 288. We compared our results with photometry published by Bolte (1992). Six bright stars common to both studies could be easily identified in our data using finding charts published by Bolte (1992). These were stars 3, 13, 23, 32, 50 and 109 from Table 5 published in Bolte's paper. For these six stars we obtained $\delta V = 0.038 \pm 0.010$ and $\delta(B - V) = -0.018 \pm 0.003$, where the differences are given in the sense JK − B.

The derived CMD of the cluster is shown in. Fig. 3. Left panel of this figure shows photometry for all measured stars while the CMD presented in the right panel was cleaned from stars of relatively poor photometry. Poor measurements were first flagged in the photometry files generated by DAOPHOT/ALLSTAR. A given measurement was considered to be poor when the formal error of photometry was 2.5 times or more larger than the average error of photometry for other stars of comparable magnitudes. Stars for which either V or B photometry was flagged as poor, were not plotted in Fig. 3b.

## 3. Results for variable stars

The search for potential short-period variables was conducted using a data base containing light curves for a total of 16323 stars. Only 7098 objects with average magnitudes lower than $V = 20.0$ and whose light curves included more than 45 points were examined for variability. To select potential variables we employed two methods which are described in details by Kaluzny

– 4 –et al. (1995). We discovered 5 new periodic variables and detected the known variable V2 (Hollingsworth and Liller 1977). The variable V1 (Oosterhoff 1943) was not included in the data base - its image was badly overexposed on all long exposed frames. The rectangular coordinates of variables V1-7 are listed in Table 2. They correspond to the positions of variables on the "template" image which was submitted to editors of A&A (see Appendix A). Table 3 lists some basic characteristic of the light curves of variables V2-7. The periods were determined using the *aov* statistic (Schwarzenberg-Czerny 1989, 1991). Our periods should be treated as preliminary, especially those for two RR Lyr stars V2 and V3. Hollingsworth and Liller (1977) estimated period of V2 at 0.679 days, based on 17 data points obtained over a period of 11 days. The mean V magnitudes from Table 3 were calculated by numerically integrating the phased light curves after converting them to an intensity scale. The listed colors are based on a single 360 seconds long exposure in the B-band (see Table 1) and two bracketing exposures in the V-band. The exposures in the V-band lasted 240 seconds and the V-band photometry was interpolated for the moment of mid-exposure of the B-band frame.

Of five newly discovered variables four are SX Phe stars and one is RR Lyr star of RRc-subtype. The phased light curves of stars V2-7 are shown in Fig. 4.

## 4. Discussion and Summary

Figure 5 shows a CMD of NGC 288 with marked positions of all the 7 variables known in this cluster. Photometry of a long-period variable V1 is based on two short exposures listed in Table 1. We examined in detail light curves of 48 candidates for blue stragglers selected based on CMD presented in Fig. 3b. Specifically, we selected 48 stars with $16.5 < V < V 18.5$ and $0.17 < (B - V) < 0.42$. This sample included 5 objects exhibiting noisy but non-periodic light curves and 3 earlier detected SX Phe stars (the fourth SX Phe star was not included in a sample as its photometry was marked as relatively poor). For the remaining 40 stars we could rule out any periodic variability with periods $P < 3$ hours and the full amplitude exceeding 0.06 magnitude. Hence, the relative frequency of SX Phe stars among NGC 288 blue stragglers is higher or equal to 3/43. Note that some candidates for blue stragglers can in fact be field stars, which would increase the estimated relative frequency of occurence of SX Phe stars.

The intensity-averaged V magnitudes of two RR Lyr variables are $<V> = 15.185$ and $<V> = 15.157$ for V2 and V3, respectively. These values are only slightly smaller than the commonly adopted estimate of the level of the zero age horizontal branch of NGC 288 – in his compilation Peterson (1993) lists V(HB) = 15.31. According to Sandage (1992) the intrinsic HB width varies with metallicity and can be as large as 0.6 mag in metal-rich clusters.

Figure 6 shows location of four newly discovered SX Phe stars on a period versus absolute magnitude diagram. The relation for the F-mode pulsators and for [Fe/H] = −1.4 is shown with the soloid line. The calibration $P - L -$ [Fe/H] was taken from Nemec et al. (1995). The dashed line shows calibration derived by McNamara (1995). According to criteria introduced by McNamara (1995) V4-6 are fundamental mode pulsators while V7 is probably a first-overtone pulsator. We adopted an apparent distance modulus of the cluster $(m - M)_V = 14.71$ (Petersen 1993) while calculating the absolute magnitudes of SX Phe stars. Examination of Fig. 6 indicates that the observed luminosities of SX Phe stars in NGC 288 are well reproduced by the calibration derived by McNamara (1995). On the other hand the calibration proposed by Nemec et al. (1994) predicts values of $M_V$ which are about 0.3 mag too low in comparison with observations.

Three candidates for hot subdwarfs can be identified in Fig. 3b. They are located at $V \approx 19$ and $(B - V) \approx -0.3$. There is a clear gap between these three blue stars and the faint end of the blue horizontal branch of NGC 288. The rectangular coordinates and photometry of candidates for hot subdwarfs are given in Table 4. Finally, we note that the distribution of stars along the blue horizontal branch of NGC 288 shows an apparent clump at $V \approx 16.0$. This feature can be potentially used to test evolutionary models of HB stars.

## 5. Appendix A

Tables containing light curves of all variables discussed in this paper as well as tables with BV photometry for 8689 stars from the surveyed field of NGC 288 are published by A&A at the Centre de Données de Strasburg, where they are available in the electronic form: See the Editorial in A&A 1993, Vol. 280, page E1. We have also submitted the V-filter "template" image of the monitored field. This image can be used for identification of all variables discussed in this paper as well as for identification of all stars for which we provide BV photometry. A table with detailed log of observations used in this paper is also available from A&A data base.

This project was supported by the Polish KBN grant 2P03D00808 to JK and by NSF grant AST 93-13620 to Bohdan Paczynski. Thanks are due to Kazik Stępień for comments on the draft version of this paper.

Table 1: List of images used for construction of the CMD of NGC 288.

| $T_{\exp}$ | Filter | Airmass | FWHM |
| --- | --- | --- | --- |
| sec | | | arcsec |
| 30 | V | 1.002 | 1.16 |
| 240 | V | 1.003 | 1.14 |
| 70 | B | 1.025 | 1.60 |
| 360 | B | 1.005 | 1.40 |

Table 2: Rectangular coordinates of NGC 288 variables. X and Y are expressed in pixels and refer to the "template" image (see Appendix A).

| Name | X | Y |
| --- | --- | --- |
| V1 | 1334.7 | 772.9 |
| V2 | 1148.3 | 874.3 |
| V3 | 1364.2 | 629.6 |
| V4 | 1276.2 | 969.7 |
| V5 | 1203.8 | 836.4 |
| V6 | 1289.6 | 992.6 |
| V7 | 1323.8 | 855.9 |

Table 3: Light-curve parameters for NGC 288 variables V2-7. $\Delta$V is the range of observed variations in the V band. The period is given in days. The (B-V) color was measured at random phases.

| Name | Type | Period | $<V>$ | $V$max | $\Delta$V | (B-V) |
| --- | --- | --- | --- | --- | --- | --- |
| V2 | RRab | 0.6811 | 15.18 | 14.56 | 1.06 | 0.38 |
| V3 | RRc | 0.3009 | 15.13 | 14.94 | 0.42 | 0.28 |
| V4 | SX | 0.07908 | 17.23 | 17.03 | 0.30 | 0.29 |
| V5 | SX | 0.05108 | 17.54 | 17.24: | 0.46: | 0.16 |
| V6 | SX | 0.06723 | 17.28 | 17.03 | 0.41 | 0.30 |
| V7 | SX | 0.04165 | 17.92 | 17.88 | 0.07 | 0.28 |

Table 4: Rectangular coordinates and BV photometry for 3 faint blue stars from the field of NGC 288. X and Y are expressed in pixels and refer to the "template" image (see Appendix A).

| Name | X | Y | V | $\sigma_V$ | B-V | $\sigma_{B-V}$ |
|---|---|---|---|---|---|---|
| b1 | 930.31 | 1292.75 | 19.15 | 0.027 | -0.266 | 0.035 |
| b2 | 1311.04 | 955.56 | 18.59 | 0.043 | -0.337 | 0.052 |
| b3 | 1455.17 | 841.50 | 19.33 | 0.021 | -0.275 | 0.026 |

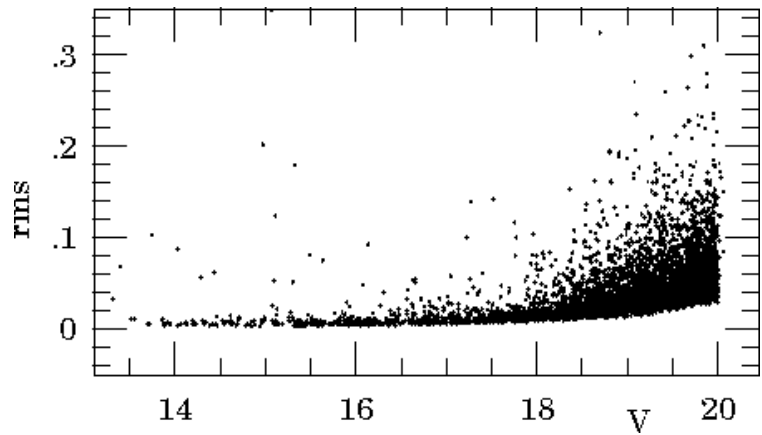

Fig. 1.— The single-measurement errors of our photometry versus the average V magnitudes for 7098 stars whose light curves were examined for variability.

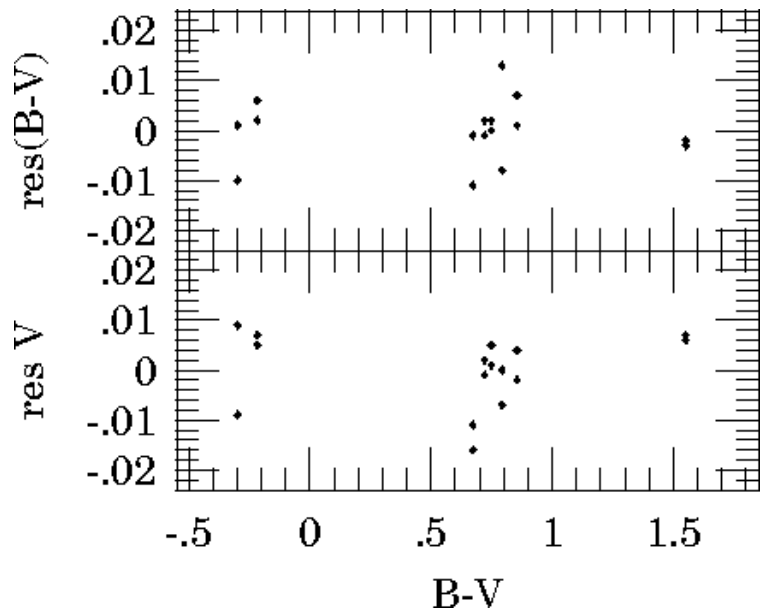

Fig. 2.— Residuals for Landolt standards observed at the end of the night of July 17/18, 1995.

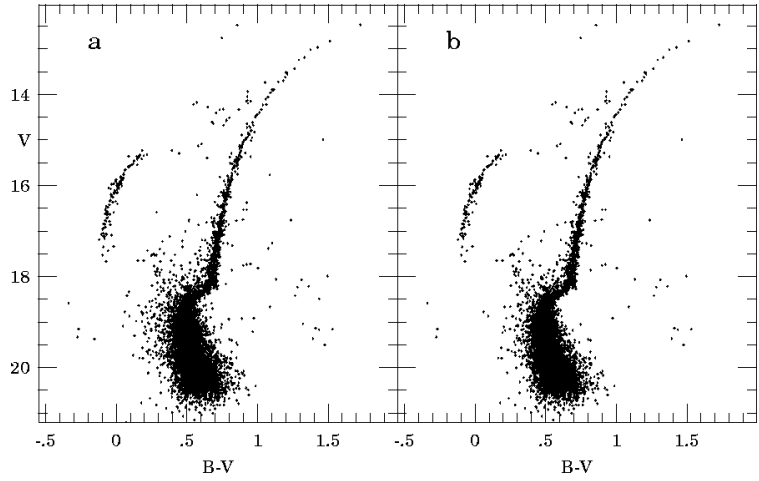

Fig. 3.— The color-magnitude diagrams for a) a sample of all stars with derived BV photometry; b) a sample of stars with removed objects of relatively poor photometry.

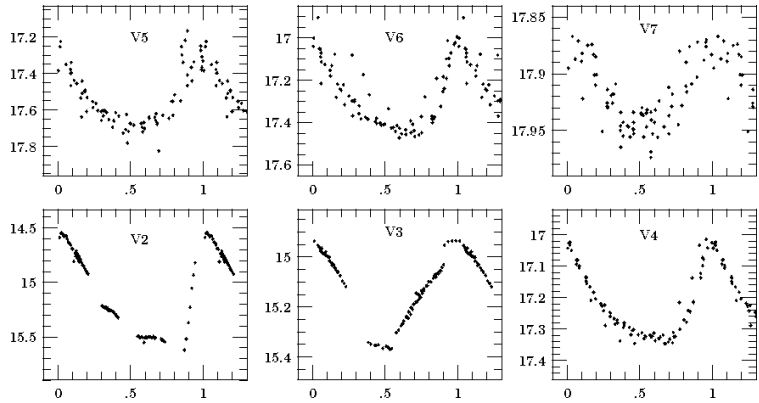

Fig. 4.— Phased V light curves for NGC 288 variables V2-7.

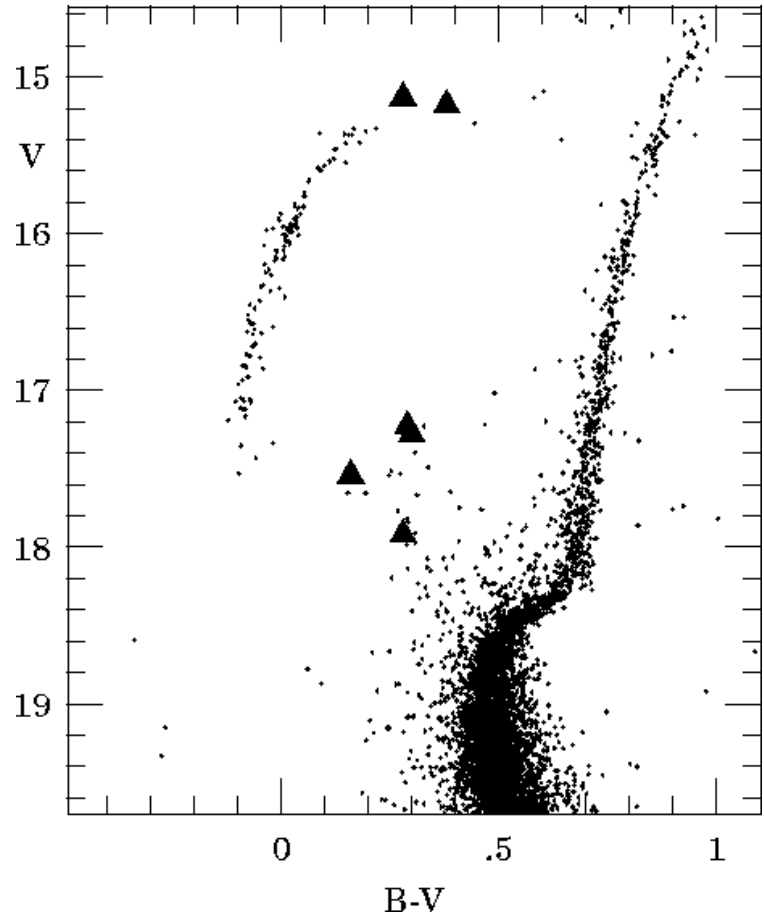

Fig. 5.— The color-magnitude diagram of NGC 288 with the positions of the variables. Stars with unreliable photometry were not plotted.

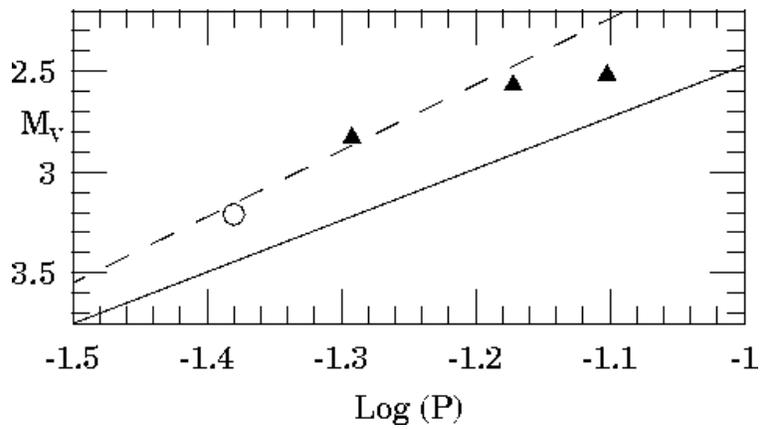

Fig. 6.— Period versus absolute magnitude diagram for SX Phe stars in NGC 288. The solid line marks a relation proposed by Nemec et al. (1994) while the dashed line shows a relation derived by McNamara (1995). The F-mode pulsators V4-6 are marked with triangles while the probable H-mode pulsator V7 is marked with an open circle.